\documentclass[10pt,journal]{IEEEtran}

\usepackage{array}
\usepackage{graphicx}   
\usepackage{float}
\usepackage{subfig}

\usepackage{amsmath}

\usepackage{amssymb}
\usepackage{cite}
\usepackage{enumerate} 

\usepackage{amsfonts} 
\usepackage{epsfig} 
\usepackage{epstopdf}
\usepackage{color}
\usepackage{eqlist}
\usepackage{pifont}
\usepackage{booktabs}   
\usepackage{multirow}   
\usepackage{makecell}   

\allowdisplaybreaks[4]  

\usepackage{indentfirst}
\usepackage{hyperref}   
\hypersetup{hypertex=true,
colorlinks=true,
linkcolor=blue,
anchorcolor=blue,
citecolor=blue}

\usepackage{fancyhdr}
\begin{document}

\title{
Self-Adjusting Prescribed Performance Control for Nonlinear Systems with Input Saturation
}
\author{Zhuwu~Shao, Yujuan~Wang$^{*}$, Huanyu Yang, Yongduan~Song,~\IEEEmembership{Fellow, IEEE}{}
        {}
\thanks{This work was supported by the National Natural Science Foundation
of China under grant (No. 61991400, No. 61991403, No. 62273064, No.62250710167, No. 61860206008, No. 61933012, and No. 62203078), and in part by the Innovation Support Program for International Students Returning to China under grant (No. cx2022016). \emph{(Corresponding author: Yujuan Wang.)}

Z. Shao, Y. Wang, H. Yang and Y. Song are with the State Key Laboratory of Power Transmission Equipment \& System Security and New Technology, and School of Automation, Chongqing University, Chongqing, 400044, China (e-mail: zwshao@cqu.edu.cn; yjwang66@cqu.edu.cn; hyyang@stu.cqu.edu.cn; ydsong@cqu.edu.cn).

}
}

\markboth{}%
{Shao, Wang, Yang and Song \MakeLowercase{\textit{et al.}}: }
\maketitle

\begin{abstract}
Among the existing works on enhancing system performance via prescribed performance functions (PPFs), the decay rates of PPFs need to be predetermined by the designer, directly affecting the convergence time of the closed-loop system.
However, if only considering accelerating the system convergence by selecting a big decay rate of the performance function, it may lead to the severe consequence of closed-loop system failure when considering the prevalent actuator saturation in practical scenarios.
To address this issue, this work proposes a control scheme that can flexibly self-adjust the convergence rates of the performance functions (PFs), aiming to achieve faster steady-state convergence while avoiding the risk of error violation beyond the PFs' envelopes, which may arise from input saturation and improper decay rate selection in traditional prescribed performance control (PPC) methods.
Specifically, a performance index function (PIF) is introduced as a reference criterion, based on which the self-adjusting rates of the PFs are designed for different cases, exhibiting the following appealing features:
1) It can eliminate the need to prespecify the initial values of the PFs.
In addition, it can also accommodate arbitrary magnitudes of initial errors while avoiding excessive initial control efforts.
2) Considering actuator saturation, this method can not only reduce the decay rates of the PFs when necessary to avoid violation of the PFs, but also increase the decay rates to accelerate system convergence when there is remaining control capacity.
Finally, several numerical simulations are conducted to confirm the effectiveness and superiority of the proposed method.

{\bf\emph{ Key words-}\rm \textbf{Adaptive control, fast convergence, input saturation, self-adjusting prescribed performance function, uncertain nonlinear systems.} }
\end{abstract}

\section{Introduction}
\IEEEPARstart{I}N recent years, driven by the pursuit of faster convergence and higher control accuracy, a number of remarkable works have emerged on accelerating convergence and prescribed performance control (PPC) \cite{bechlioulis2008robust,mehdifar2020prescribed,song2017time,zhao2021adaptive,song2016adaptive}.
In this field of PPC, there are currently two mainstream strategies, namely funnel control (FC) \cite{ilchmann2008high,hopfe2010funnel,chowdhury2019funnel} and prescribed performance bound (PPB) based control \cite{bechlioulis2008robust,bai2024unified,wang2022prescribed,wang2024fixed}.
Both methods introduce funnel functions as enveloping bounds for the system errors to ensure that the tracking errors remain confined within their envelopes throughout the entire system operation.
Furthermore, \cite{zhang2017prescribed} and \cite{zhao2019tracking} have also achieved PPC for nonlinear systems by constructing novel error transformation functions and implementing a new PPB framework.
Notably, \cite{zhao2021adaptive} has further accomplished global PPC based on the PPB method through 
the normalized transformations.
Specifically, a set of PFs with initial values tending to infinity is utilized, ensuring that the initial values of the system error always remain within the bounds defined by the PFs.
Moreover, \cite{zhao2023unifying} and \cite{zhang2024global} have extended the relevant results to more general nonlinear systems. 

Although both FC and PPB have proven effective in implementing performance specifications, it should be noted that the appropriate selection of the decay rate of the PF is crucial for achieving the desired tracking performance, regardless of which PPC scheme is employed.
On one hand, if a small decay rate is directly set, such methods, especially in terms of transient performance with small initial errors, have very small effects compared to methods without PPC.
On the other hand, if a large rate is set, particularly when the desired tracking error needs to converge to a predetermined tight set within a predefined time, the required peak input by the controller will rapidly increase as the initial error grows.
However, in practice, the presence of physical input saturation imposes constraints on the magnitude of the control signal \cite{wen2011robust}.
In such scenarios, maintaining the original convergence rate of the PFs could potentially result in instability within the closed-loop system \cite{yong2020flexible, trakas2023robust, fotiadis2023input}.
Regarding the above-mentioned issues, a solution was proposed within the FC framework \cite{hopfe2010funnel}.
Specifically, a feasibility condition was introduced, targeting the lower bound of input saturation.
This condition ensures that the control system can successfully achieve the control objective while maintaining boundedness of both the state and gain function. 
It is worth highlighting that the majority of existing literature that simultaneously considers tracking performance constraints and input constraints tends to operate under the assumption that this feasibility condition holds, which is, in fact, relatively conservative.
More recently, within the PPB framework, by introducing an innovative auxiliary system, a novel algorithm was developed in \cite{yong2020flexible} to relax the performance bounds online when input saturation occurs.
Subsequently, such methods were extended to many types of nonlinear systems \cite{ji2022saturation1,ji2022saturation2,xie2023flexible}.
Conversely, another type of algorithm was proposed in \cite{fotiadis2023input} and \cite{fotiadis2023inputC} to modify the reference signal online when input saturation occurs.
This advancement also enables the feasibility of the PPC method under input saturation constraints.
However, the aforementioned studies can currently only relax the PF when input saturation occurs and gradually restore it as saturation subsides.
They lack the capability to actively contract the PF in an attempt to achieve better performance.
Moreover, to the best of our knowledge, there are still relatively few research achievements in this area currently.

To tackle the aforementioned issues, this work presents a straightforward self-tuning method for the decay rate of the PF.
In contrast to previous studies \cite{yong2020flexible, trakas2023robust, fotiadis2023input, ji2022saturation1, ji2022saturation2, xie2023flexible} that relied on the saturation deficiency term to design auxiliary increments or modify the reference signal, a performance index function (PIF) is introduced.
In this context, the PIF, originally used for monitoring actuator faults in \cite{ouyang2020adaptive} and \cite{ouyang2017adaptive}, is utilized as a baseline for adjusting the decay rate of the PF.
Furthermore, through the output-based performance feedback, not only is the PF capable of being relaxed when input saturation occurs, but it can also be further tightened when a significant margin in the input is detected, which distinguishes it from current related works.
Ultimately, the PF is allowed to adjust its own decay rate online based on the overall system performance. This is a more convenient and flexible approach compared to previous approaches that utilized fixed exponential decay rates.


\textbf{Notation:}
Throughout this paper, $\Re$ denotes the set of real numbers.
$\|x\|$ denotes the Euclidean norm of vector $x$.
$\lambda_{\text{min}}(\Gamma)$ and $\lambda_{\text{max}}(\Gamma)$ denote the minimum and maximum eigenvalue of matrix $\Gamma$, respectively.
For convenience, some/all arguments of certain functions/variables will be omitted or replaced by $(\cdot)$ if no confusion is likely to occur.

\section{Problem Statement}
Consider the following strict-feedback nonlinear systems,
\begin{equation}\label{system}
    \left\{
        \begin{aligned}
            &\dot{x}_{k} = x_{k+1} + \theta^T \varphi_k(\bar{x}_i) + d_k(t), \ k=1,\cdots,n-1\\
            &\dot{x}_{n} = u + \theta^T \varphi_n(\bar{x}_n) + d_n(t),\\
            &y = {{x}_{1}},
        \end{aligned}
    \right.
\end{equation}
where $x_i \in \Re$ ($i = 1, \cdots, n$) is the system state and is assumed available for measurement,
with $\bar{x}_i$ = $[x_{1}, \cdots, x_{i}]^{T}$,
$\theta \in \Re^r$ and $\phi_i(\bar{x}_i) \in \Re^r$ are unknown constant vector and known smooth nonlinear functions,  respectively,
$d_i(t)$ denote unmeasured disturbances satisfying $|d_i(t)| \leq \bar{d}_i$ with $\bar{d}_i$ an unknown constant,
$y \in \Re$ is the system output, and $u \in \Re$ is the actual control input exhibits the following symmetric saturation nonlinearity,
\begin{equation}\label{input_sat}
    u(v) = \text{Sat}(v) =
    \left\{
        \begin{aligned}
            & \bar{u} \cdot \text{sign}(v),  &\  |v| > \bar{u}, \\
            & v,        &  |v| \leq \bar{u},
        \end{aligned}
    \right.
\end{equation}
where $v$ denotes the desired control input, $\bar{u}$ represents a known physical limitation of $u$.

To deal with the sharp corners of $\text{Sat}(v)$, we split $u$ into the following two parts:
\begin{align}\label{Sat_u}
    & u(v) = \text{Sat}(v) = h(v) + g(v),
\end{align}
where $h(v) = \bar{u} \, \text{tanh}(v/\bar{u})$, $g(v) = \text{Sat}(v) - h(v)$ with
$|g(v)| \leq \bar{u}(1 - \text{tanh}(1)) \leq 0.24\bar{u} = \bar{g}$.

Define $e = x_1 - y_d$ as the tracking error, where $y_d$ represents the reference signal.
The PF for $e$, which is independent of the initial conditions \cite{zhao2021adaptive}, is given by:
\begin{align}\label{PPF}
P(t) := I(\Psi) = \frac{\sqrt{1-\Psi^2_{\infty}}\Psi}{\sqrt{1-\Psi^2}},
\end{align}
where $\Psi(t)$ is defined as:
\begin{align}\label{varPsi}
\Psi(t) = (1 - \Psi_{\infty})\exp^{-\delta t} + \Psi_{\infty}.
\end{align}
In the above equations, $0 < \Psi_{\infty} \ll 1$ represents the desired steady-state control accuracy;
$\delta(t) > 0$ for all $t > 0$, with the initial value $\delta(0)=0$, serves as the time-varying exponential decay rate of $\Psi(t)$ and indirectly determines the actual decay rate of the PF ($\dot{P}(t)$).
For clarity and convenience in expression, the decay rate of the PF will be directly denoted as $\delta(t)$ and designed in subsequent sections.
The control objective is to design a self-tuning method for the decay rate of the PF within the framework of PPB-based control, specifically for the nonlinear system (\ref{system}) under input saturation constraints.
The aim is to enable the system output $y(t)$ to closely and rapidly track the desired trajectory $y_d(t)$ within the permissible range of inputs. 
Specifically, when PIF detects that the decay rate of the tracking error cannot reach the current decay rate of the PF under the given input constraints, $\delta(t)$ will adaptively decrease.
Conversely, $\delta(t)$ will moderately increase when the decay rate of the tracking error can exceed the current decay rate of the PF.

To this end, the following Assumptions are imposed.

\emph{Assumption 1:}\label{assumption1} The desired trajectory $y_d$ and its derivatives up to $n$th are known, bounded, and piecewise continuous.

\emph{Assumption 2:}\label{assumption2} The unknown parameter vector $\theta$ lies in a known bounded convex set
\begin{align}\label{theta}
    &\Pi_{\theta} = \{ \hat{\theta} \in \Re^r | \mathcal {P}(\hat{\theta}) \leq 0 \}
\end{align}
where $\mathcal {P}$ is a convex smooth function and $\hat{\theta}$ is the estimate of $\theta$.
Naturally, an upper bound of $\|\theta\|$, denoted by $\theta_M$, is known, such that $\|\theta\| \leq \theta_M$.

\emph{Assumption 3:}\label{assumption3} The plant is input-to-state stable (ISS).

\emph{Remark 1:} As illustrated in an example provided in \cite{wen2011robust}, \hyperref[assumption3]{Assumption 3} is reasonable since an unstable device cannot be globally stable in the presence of input saturation.
Moreover, this further implies that, if the controller design is appropriate, there exists a sufficiently small positive constant $\underline{\delta}_0$ such that $|e(t)| < P(t)|_{\delta = \underline{\delta}_0}$ for all $t \geq 0$.

\section{Control Design without Considering Input Constraints}
\subsection{Performance Index Function}
First, we shall design the controller using the backstepping technique without considering the input saturation limitation and the external disturbances.


\begin{table}[tp]
    \small
    \centering
    \caption{Controller Design under the Backstepping Approach}\label{table1}
    \label{tab:univ-compa}
    \resizebox{\linewidth}{!}{
    \begin{tabular}{llr}
    \toprule

        \emph{Errors:} \\
        \multirow{2}*{\makecell[l]{ $z_1 = s = \frac{\zeta}{1 - \zeta^2} := S(\zeta)$, $\zeta \in (-1, 1)$,}}  \\ \\
        $z_i = x_i - \alpha_{i-1}$, $i = 2, \cdots, n$. \\ \midrule

        \multirow{1}*{\makecell[l]{ \emph{Step 1:} }} \\
        \multirow{2}*{\makecell[l]{ $V_1 = \frac{1}{2}z^2_1 + \frac{1}{2} \tilde{\theta}^T \Gamma^{-1} \tilde{\theta}$, }}   \\ \\
        \multirow{2}*{\makecell[l]{ $\alpha_1 = -\frac{1}{\mu_1}(c_1 z_1 + \mu_2) - \hat{\theta}^T\varphi_1 + \dot{y}_d$. }} \\ \\ \midrule

        \multirow{1}*{\makecell[l]{ \emph{Step 2:} }} \\
        \multirow{2}*{\makecell[l]{ $V_2 = V_1 + \frac{1}{2}z^2_2$, }}  \\ \\
        \multirow{3}*{\makecell[l]{ $\alpha_2 = -c_2 z_2 - \mu_1 z_1 - \hat{\theta}^T\omega_2 + {\frac{\partial \alpha_{1}}{\partial x_1} x_{2}}$ \\ $\quad \quad \ \, + \sum^{1}_{k=0} {\frac{\partial \alpha_{1}}{\partial y^{(k)}_d} y^{(k+1)}_d + \sum^{1}_{k=0} \frac{\partial \alpha_{1}}{\partial \beta^{(k)}} \beta^{(k+1)}} + \frac{\partial \alpha_{1}}{\partial \hat{\theta}} {\Gamma \tau_2}$. }}   \\ \\ \\ \midrule

        \multirow{1}*{\makecell[l]{ \emph{Step i ($i = 3, \cdots, n$):} }} \\
        \multirow{2}*{\makecell[l]{ $V_i = V_{i-1} + \frac{1}{2}z^2_i$, }} \\ \\
        \multirow{4}*{\makecell[l]{ $\alpha_i = -c_i z_i - z_{i-1} - \hat{\theta}^T\omega_i + \sum^{i-1}_{k=1} {\frac{\partial \alpha_{i-1}}{\partial x_k} x_{k+1}} + \frac{\partial \alpha_{i-1}}{\partial \hat{\theta}} {\Gamma \tau_i}$ \\ $\quad \quad \ \, + \sum^{i-1}_{k=0} \left[ {\frac{\partial \alpha_{i-1}}{\partial y^{(k)}_d} y^{(k+1)}_d + \frac{\partial \alpha_{i-1}}{\partial \beta^{(k)}} \beta^{(k+1)}} \right] + \sum^{i-1}_{k=2} {\frac{\partial \alpha_{k-1}}{\partial \hat{\theta}} \omega_i z_k \Gamma}$. }} \\ \\ \\ \\ \midrule

        \emph{Parameter update law:}\\
        \multirow{2}*{\makecell[l]{ $\dot{\hat{\theta}} = \text{Proj}_{\Pi_{\theta}} \{\Gamma \tau_n\}$. }} \\ \\

        \emph{Tuning functions:}\\
        \multirow{3}*{\makecell[l]{  $\tau_i = \sum^{i}_{k=1} \omega_i z_i$, $i = 1, \cdots, n$. \\
        $\omega_1 = \mu_1 \varphi_1$, \ $\omega_i = \varphi_i - \sum_{k=1}^{i-1} \frac{\partial \alpha_{i-1}}{\partial x_k} \varphi_k$, $i = 2, \cdots, n$. }} \\ \\ \\

    \bottomrule
    \end{tabular}}
\end{table}

Based on Lyapunov stability analysis technique, the backstepping design procedure is summarized in \hyperref[table1]{Table I}, where $\zeta = \beta(t) \eta(e)$, $\beta(t) = 1/\Psi(t)$, $\eta(e) = \frac{e}{\sqrt{e^2 + 1 - \Psi^2_{\infty}}}$,
$\mu_1 = \mu \beta \rho$, $\mu_2 = \mu \dot{\beta} \eta$, $\mu = \frac{1+\zeta^2}{(1-\zeta^2)^2}$,
$\tilde{\theta} = \theta - \hat{\theta}$, $\Gamma$ is a positive definite design matrix, $c_i$, $i = 1, \cdots, n$, is a  positive design parameter.
Therefore, for the plant (\ref{system}) without considering the input saturation, under Assumptions \hyperref[assumption1]{1} and \hyperref[assumption2]{2}, by employing the control law and parameter update law as
\begin{align}\label{u-1}
    u = &\ \alpha_n,    \\
    \dot{\hat{\theta}} =&\ \text{Proj}_{\Pi_{\theta}} \{\Gamma \tau_n\},    \label{dtheta-1}
\end{align}
it can be concluded that
\begin{align}\label{dV_n-1}
    &\dot{V}_n \leq -\sum^{n}_{k=1} c_k z^2_{k} \leq 0,
\end{align}
where $\text{Proj}(\cdot)$ is the projection operator given in \cite{krstic1995nonlinear}, which ensures that $\|\hat{\theta}\| \leq \theta_M$.

From (\ref{dV_n-1}), the definition of $V_n$ and \hyperref[assumption2]{Assumption 2}, we have
\begin{align}\label{V_n-1}
    V_n(t) &\leq V_n(0) \nonumber \\
    &\leq \frac{1}{2}\bigg{[} \sum^n_{k=1} z^2_k(0) + \lambda_{\text{max}}(\Gamma^{-1})(\theta_M + \| \hat{\theta}(0) \|)^2 \bigg{]} < \frac{1}{2} \mu_0^2,
\end{align}
where
\begin{align}
    \mu_0
    &= \bigg{[} \sum^n_{k=1} z^2_k(0) + \lambda_{\text{max}}(\Gamma^{-1})(\theta_M + \| \hat{\theta}(0) \|)^2 + \varepsilon_0 \bigg{]}^{\frac{1}{2}},   \nonumber
\end{align}
with $0 < \varepsilon_0 \ll 1$ is a constant.

Based on the analysis above, we define the following PIF:
\begin{align}\label{PIF}
    p(t) = I[\Psi(t)S^{-1}(\mu_0)].
\end{align}
The relevant results are summarized in the following theorem.

\emph{Theorem 1:} \label{Theorem 1}
For the plant (\ref{system}) without considering the input saturation and external disturbances, under Assumptions \hyperref[assumption1]{1} and \hyperref[assumption2]{2}, by employing the control law (\ref{u-1}) and parameter update law (\ref{dtheta-1}), the following relationship is guaranteed to hold for $\forall t \geq 0$:
\begin{align}\label{e_bound}
    -P(t) < -p(t) < e(t) < p(t) < P(t).
\end{align}

\emph{Proof:} From (\ref{V_n-1}) and the definition of $V_n$, we obtain
\begin{align}
    \frac{1}{2}s^2 < \frac{1}{2}\mu_0^2.
\end{align}
Therefore, we can conclude that $-\mu_0 < s < \mu_0$ for all $t \geq 0$.
Since $\frac{\partial s}{\partial \zeta} = \mu > 0$, which implies that $S(\zeta)$ is strictly monotonically increasing with respect to $\zeta$ over the interval $(-1, 1)$.
Thus, its inverse function $S^{-1}(s)$ is also strictly monotonically increasing with respect to $s$, indicating that $S^{-1}(-\mu_0) < S^{-1}(s) < S^{-1}(\mu_0)$.
Furthermore, noting that $S^{-1}(s) = \zeta = \beta(t) \eta(e)$, with $\beta(t) = 1/\Psi(t) > 0$, we can deduce:
\begin{align}
    \Psi(t)S^{-1}(\mu_0) < \eta(e) < \Psi(t)S^{-1}(\mu_0).
\end{align}
Moreover, it follows from $s = \frac{\zeta}{1 - \zeta^2} := S(\zeta)$ that
\begin{align}
    S^{-1}(\mu_0) = \frac{\sqrt{4\mu_0^2+1}-1}{2\mu_0} < \frac{\sqrt{4\mu_0^2}}{2\mu_0} = 1.
\end{align}
Then, one can obtain
\begin{align}
    \Psi(t) < \Psi(t)S^{-1}(\mu_0) < \eta(e) < \Psi(t)S^{-1}(\mu_0) < \Psi(t).
\end{align}
According to Lemma 1 in \cite{zhao2021adaptive}, it can be concluded that the function $I(\cdot)$ is strictly monotonically increasing over the interval $(-1, 1)$.
Additionally, noting that $e(t) = I[\eta(e)]$, we can therefore conclude that (\ref{e_bound}) holds for $\forall t \geq 0$.
$\hfill \blacksquare$

\emph{Remark 2:}
It is worth noting that the PIF designed in this work is inspired by the monitoring function in \cite{ouyang2020adaptive} and \cite{ouyang2017adaptive};
however, it should be recognized that due to their different roles in respective tasks, the monitoring function in \cite{ouyang2020adaptive} and \cite{ouyang2017adaptive} must be strictly respected under normal circumstances.
In contrast, the PIF in our work serves merely as a performance metric that can be violated.
The emphasis of our work lies in flexibly adjusting the PFs guided by this metric to achieve better system performance.
Therefore, the effect of unmeasured disturbances is considered in our system model from the outset, and the PIF design does not need to account for disturbances, fundamentally differing from the monitoring function in \cite{ouyang2020adaptive} and \cite{ouyang2017adaptive}.
Additionally, it is worth clarifying that although this work introduces global PFs based on the conclusions in \cite{zhao2021adaptive}, our method is not limited to such global PFs.
By employing our proposed method, there is no need to worry about large initial PF values undermining the effectiveness of traditional PPC methods.
This is because even with large initial PF values, our method will promptly drive the PFs to shrink rapidly from a large envelope to a smaller one that the current input can accommodate.


\subsection{Self-Tuning Decay Rate Design without Input Saturation Consideration}
In theory, since $P(0) \to \infty$ and the input can be sufficiently large, the decay rate $\delta(t)$ can be arbitrarily set for any arbitrarily large initial error $e(0)$.
However, it should be noted that $\dot{P}(0) \to -\infty$ as $\delta(t) \neq 0$, which means that the actual decay rate of the PF is extremely large within a small interval near the initial time moment.
Due to the inherent challenges, it becomes difficult to achieve true global convergence in practice for the global PPC method proposed in \cite{zhao2021adaptive}, which considers a constant decay rate, even in the absence of input saturation constraints.

Given this, to attain genuine global PPC, we introduce the following self-tuning law of the decay rate of the PF:
\begin{equation}\label{d_delta_p}
    \dot{\delta} =
    \left\{
        \begin{aligned}
            & 0, & \text{if} \ p \leq \Psi_\infty,    \\
            & \varrho_1 \left(\frac{|e|}{p} - 1\right)^2 := \Delta_1\left(\frac{|e|}{p}\right),  &  \text{otherwise}, \\
        \end{aligned}
    \right.
\end{equation}
with the initial value $\delta(0) = 0$ and $\dot{\delta}(0) = 0$, where $\varrho_1$ is a positive design parameter.

Based on \hyperref[Theorem 1]{Theorem 1}, it is evident that when there are no external disturbances or input saturation constraints, the output error of the system will always remain within the bounds of PIF.
Consequently, it will consistently satisfy the condition $|e(t)| < p(t)$.
Therefore, from (\ref{d_delta_p}), it can be seen that $\Delta_1(\cdot)$ exhibits the following easily verifiable properties:
\begin{enumerate}
  \item $\Delta_1(\cdot) \geq 0$ is of $n$-th order differentiability for all $|e| \leq p$;
  \item $\Delta_1(\cdot)$ is strictly monotonically decreasing on the internal $(0,1]$, with $\Delta_1(0) = \varrho_1$ and $\Delta_1(1) = 0$;
  \item $\dot{\Delta}_1(\cdot)$ is strictly monotonically increasing on the internal $(0,1]$, with $\dot{\Delta}_1(1) = 0$.
\end{enumerate}

\begin{figure}[tp]\label{fig1}
\centering
\setlength{\belowcaptionskip}{0mm}
\includegraphics[scale=0.64]{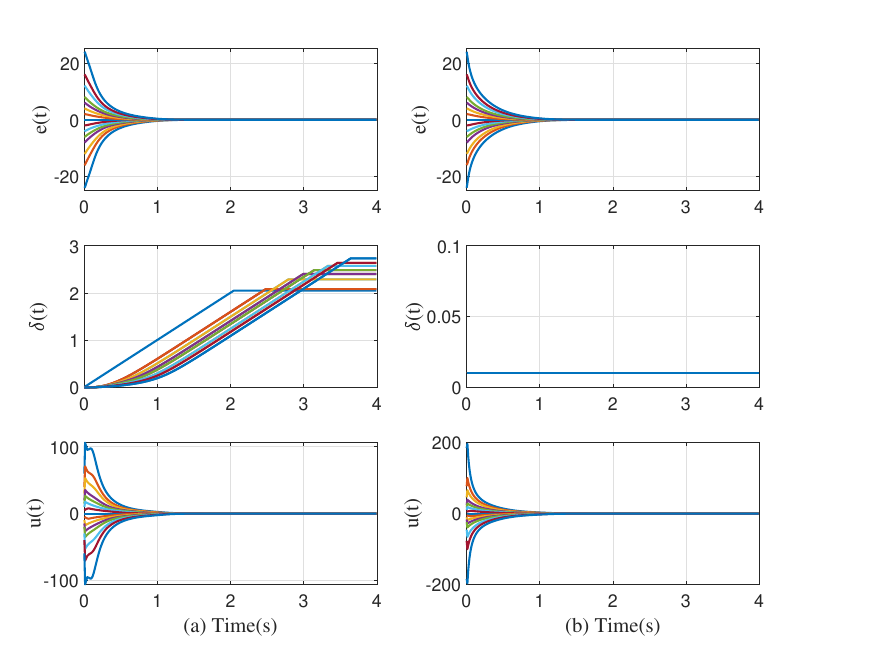}
\captionsetup{font={small}}
\caption{The control performance corresponding to (a) this work, and (b) work \cite{zhao2021adaptive}.}
\setlength{\abovecaptionskip}{1cm}   
\end{figure}

To validate the effectiveness of the proposed self-tuning decay rate of the PF, we first consider the $1$st-order form of system (\ref{system}) without the disturbance $d_1$ and input constraint limitations, i.e.,
\begin{equation}\label{1-order system}
    \begin{aligned}
        &\dot{x}_{1} = u + \theta^T \varphi_1,
    \end{aligned}
\end{equation}
with $\theta = 1$, $\theta_M = 2$ and $\varphi_1 = x_1$.
Moreover, we compare it with the classical global PPC method presented in \cite{zhao2021adaptive} under different initial error conditions.
The desired signal is defined as $y_d = 0$.
For the shared control parameters, we choose $c = 5$, $\Gamma = 1$, $\Psi_\infty = 0.05$, and initialize $\hat{\theta}(0) = 0$.
To obtain similar results, we set the parameter $\delta$ in \cite{zhao2021adaptive} as $\delta = 0.01$, and in our work, we set $\varrho_1 = 1$ and $\varepsilon_0 = 0$.
The simulation results are shown in \hyperref[fig1]{Fig. 1}.
It can be observed that although both methods can achieve similar control performance for different initial errors by adjusting the parameters, our improvement effectively reduces the required input peak values as stated in \cite{zhao2021adaptive}.
Similarly, if the parameter $\delta$ in \cite{zhao2021adaptive} is increased, the corresponding input amplitudes will also increase significantly, which is actually very unfavorable for practical engineering applications, and can be effectively mitigated in this work.

\section{Self-Tuning Decay Rate and Controller Design under Input Saturation}
\subsection{Self-Tuning of Decay Rate}
Based on the analysis in the previous section, especially the simulation verification, it can be observed that with a certain set of initial errors, if the decay rate in \cite{zhao2021adaptive} is set small enough, the global PPC method can still achieve fairly good results.
However, for a small constant decay rate $\delta_c > 0$, its transient performance is actually not superior, especially when the initial error of the system is also small.
Moreover, once the inevitable input saturation and external disturbances in practical systems are considered, the system output error, especially its transient behavior, may not meet the predefined convergence rate requirements,
which can lead to singularity in the function $S(\cdot)$ and consequently fail traditional PPC methods.
Therefore, to address the above issues, the following update rate is introduced:
\begin{equation}\label{d_delta}
    \dot{\delta} =
    \left\{
        \begin{aligned}
            & \varrho_1 \left(\frac{|e|}{p} - 1\right)^2 := \Delta_1\left(\frac{|e|}{p}\right),        &  |e| \leq p, \\
            & \varrho_2 \ \text{log}\bigg{(} \frac{p - |e|}{P-p} + 1 \bigg{)} \delta := \Delta_2(|e|)\delta,  &\, |e| > p,
        \end{aligned}
    \right.
\end{equation}
with $\dot{\delta}(t) = 0$ as $p(t) \leq \Psi_\infty$ and the initial value condition $\delta(0) = 0$ and $\dot{\delta}(0) = 0$,
where $\varrho_1$ and $\varrho_2$ are two positive design parameters, $\text{log}$($\cdot$) denotes the natural logarithmic function. 
Moreover, $\Delta_2(\cdot)$ exhibits the following verifiable properties:
\begin{enumerate}
  \item $\Delta_2(\cdot) < 0$ is of $n$-th order differentiability for all $|e| \in (p,P)$;
  \item $\Delta_2(p) = 0$, $\mathop {\lim }_{|e| \to {P^ - }} \Delta_2 \to  - \infty$;
  \item $\Delta_2(\cdot)$ and its derivative $\dot{\Delta}_2(\cdot)$ are both monotonically decreasing on the internal $(p,P)$, with $\dot{\Delta}_2(p) \neq 0$.
\end{enumerate}
It should be noted that (\ref{d_delta}) is a specific function and can be replaced by other functions that satisfy the above properties.

In fact, the auxiliary system (\ref{d_delta}) establishes a connection between the input and the performance constraints.
It uses the PIF (\ref{PIF}) as a reference.
If the system error is within the envelope of the PIF and sufficiently far from this envelope, it indicates that the current input still has potential to further improve the error decay rate.
On the contrary, if the system error exceeds the envelope of the PIF, it means that the input cannot satisfy the current decay rate of the PF.
Therefore, it is necessary to immediately reduce $\delta(t)$ to avoid singularity in $S(\cdot)$.
Furthermore, the greater the relative distance between the system error and the envelope of the PIF, the faster the change in $\delta(t)$ is required, especially when $|e(t)| \to P^-(t)$, $\Delta_2 \to -\infty$.
Combining this with \hyperref[assumption3]{Assumption 3}, if the controller design is appropriate, it is sufficient to ensure that $|e(t)| < P(t)$ for all $t \geq 0$. 
Additionally, it is important to note that these two cases mentioned above fundamentally have different impacts on the closed-loop control system.
The former may cause a temporary decrease in transient performance, while the latter may lead to the paralysis of the system.
Therefore, the properties of the designed update rate function are different depending on these two distinct situations.
Specifically, although $\Delta_1(1) = \Delta_2(p) = 0$, $\dot{\Delta}_1(1) = 0$, while $\dot{\Delta}_2(p) \neq 0$. Moreover, the boundedness of $\delta(t)$ is presented in the following theorem.

\emph{Theorem 2:} Considers the plant (\ref{system}) with input saturation (\ref{input_sat}) and designs a PPC under Assumptions \hyperref[assumption1]{1-3}.
For any initial state $x_n(0)$, if the update rate (\ref{d_delta}) is adopted, there exists a positive constant $\bar{\delta}$ such that $0 < \delta(t) \leq \bar{\delta} < \infty$ for all $t > 0$.

\emph{Proof:}
Firstly, since $\delta(0) = 0$ and $|e(0)| < p(0)$, then $\dot{\delta}(0) > 0$, there exists a sufficiently small time point $T_1$ such that $\delta(t) > 0$ holds on the interval $(0, T_1)$.
Next, if $|e| > p$ holds thereafter, we have $\dot{\delta}(t) = \Delta_2\delta(t)$ with $\Delta_2 < 0$.
By integrating both sides of the above equation, we obtain $\delta(t) = \delta(T_1) \text{exp}^{-|\Delta_2| (t-T_1)}$, which implies that $\delta(t) > 0$ still holds.
On the other hand, if $|e| \leq p$ holds, especially if it holds consistently thereafter, according to \hyperref[Theorem 1]{Theorem 1}, it can be observed that there must exist a time point $T_2$ such that $p(t) \leq \Phi_\infty$ holds for any $t \geq T_2$.
As a result, there exists a positive constant $\bar{\delta}$ such that $0 < \delta(t) \leq \bar{\delta} < \infty$ for all $t > 0$.

$\hfill \blacksquare$

\emph{Remark 2:}\label{Remark 3}
The update rate (\ref{d_delta}) ensures that $\dot{\delta}(t) > 0$ for all $t > 0$.
It means that the overall trend of the tracking error is always decaying, only the rate of decay is changing.
The final steady-state accuracy is not relaxed as the decay rate of the PF decreases.

\subsection{Adaptive Control Design with Input Saturation}

\begin{table}[tp]
    \small
    \centering
    \caption{Controller Design with Input Saturation}\label{table2}
    \label{tab:univ-compa}
    \resizebox{\linewidth}{!}{
    \begin{tabular}{llr}
    \toprule

        \emph{Errors:} \\
        \multirow{2}*{\makecell[l]{ $z_1 = s = \frac{\zeta}{1 - \zeta^2} := S(\zeta)$, $\zeta \in (-1, 1)$,}}  \\ \\
        $z_i = x_i - \alpha_{i-1}$, $i = 2, \cdots, n-1$. \\
        $z_n = x_n - \alpha_{n-1} - \xi$, $i = 2, \cdots, n-1$. \\ \midrule

        \multirow{1}*{\makecell[l]{ \emph{Step 1:} }} \\
        \multirow{2}*{\makecell[l]{ $V_1 = \frac{1}{2}z^2_1 + \frac{1}{2} \tilde{\theta}^T \Gamma^{-1} \tilde{\theta}$, }}   \\ \\
        \multirow{2}*{\makecell[l]{ $\alpha_1 = -\frac{1}{\mu_1}(c_1 z_1 + \mu_2) -\mu_1 z_1 - \hat{\theta}^T\varphi_1 + \dot{y}_d$. }} \\ \\ \midrule

        \multirow{1}*{\makecell[l]{ \emph{Step 2:} }} \\
        \multirow{2}*{\makecell[l]{ $V_2 = V_1 + \frac{1}{2}z^2_2$, }}  \\ \\
        \multirow{3}*{\makecell[l]{ $\alpha_2 = -c_2 z_2 - \mu_1 z_1 - \hat{\theta}^T\omega_2 + {\frac{\partial \alpha_{1}}{\partial x_1} x_{2}} - \big{[} \big{(} \frac{\partial \alpha_1}{\partial x_1} \big{)}^2 + 1 \big{]}z_2$ \\
        $\quad \quad \ \, + \sum^{1}_{k=0} {\frac{\partial \alpha_{1}}{\partial y^{(k)}_d} y^{(k+1)}_d + \sum^{1}_{k=0} \frac{\partial \alpha_{1}}{\partial \beta^{(k)}} \beta^{(k+1)}} + \frac{\partial \alpha_{1}}{\partial \hat{\theta}} {\Gamma \tau_2}$. }}   \\ \\ \\ \midrule

        \multirow{1}*{\makecell[l]{ \emph{Step i ($i = 3, \cdots, n-1$):} }} \\
        \multirow{2}*{\makecell[l]{ $V_i = V_{i-1} + \frac{1}{2}z^2_i$, }} \\ \\
        \multirow{5}*{\makecell[l]{ $\alpha_i = -c_i z_i - z_{i-1} - \hat{\theta}^T\omega_i + \sum^{i-1}_{k=1} {\frac{\partial \alpha_{i-1}}{\partial x_k} x_{k+1}} + \frac{\partial \alpha_{i-1}}{\partial \hat{\theta}} {\Gamma \tau_i}$ \\
        $\quad \quad \ \, + \sum^{i-1}_{k=0} \Big{[} {\frac{\partial \alpha_{i-1}}{\partial y^{(k)}_d} y^{(k+1)}_d + \frac{\partial \alpha_{i-1}}{\partial \beta^{(k)}} \beta^{(k+1)}} \Big{]} + \sum^{i-1}_{k=2} {\frac{\partial \alpha_{k-1}}{\partial \hat{\theta}} \omega_i z_k \Gamma}$ \\
        $\quad \quad \ \, - \Big{[} \sum^{i-1}_{k=1}\big{(} \frac{\partial \alpha_{i-1}}{\partial x_k} \big{)}^2 + 1 \Big{]}z_i $. }} \\ \\ \\ \\ \\ \midrule

        \multirow{1}*{\makecell[l]{ \emph{Step n:} }} \\
        \multirow{2}*{\makecell[l]{ $V_n = V_{n-1} + \frac{1}{2}z^2_n$, }}  \\ \\
        \multirow{5}*{\makecell[l]{ $v = -c_n z_n - z_{n-1} -\lambda \xi - \hat{\theta}^T\omega_n + \sum^{n-1}_{k=1} {\frac{\partial \alpha_{n-1}}{\partial x_k} x_{k+1}} + \frac{\partial \alpha_{n-1}}{\partial \hat{\theta}} {\Gamma \tau_n}$ \\
        $\quad \quad \ \, + \sum^{n-1}_{k=0} \Big{[} {\frac{\partial \alpha_{n-1}}{\partial y^{(k)}_d} y^{(k+1)}_d + \frac{\partial \alpha_{n-1}}{\partial \beta^{(k)}} \beta^{(k+1)}} \Big{]} + \sum^{n-1}_{k=2} {\frac{\partial \alpha_{k-1}}{\partial \hat{\theta}} \omega_n z_k \Gamma}$ \\
        $\quad \quad \ \, - \Big{[} \sum^{n-1}_{k=1}\big{(} \frac{\partial \alpha_{n-1}}{\partial x_k} \big{)}^2 + 1 \Big{]}z_n $. }} \\ \\ \\ \\ \\\midrule

        \emph{Parameter update law:}\\
        \multirow{2}*{\makecell[l]{ $\dot{\hat{\theta}} = \text{Proj}_{\Pi_{\theta}} \{\Gamma \tau_n\}$. }} \\ \\

        \emph{Tuning functions:}\\
        \multirow{3}*{\makecell[l]{  $\tau_i = \sum^{i}_{k=1} \omega_i z_i$, $i = 1, \cdots, n$. \\
        $\omega_1 = \mu_1 \varphi_1$, \ $\omega_i = \varphi_i - \sum_{k=1}^{i-1} \frac{\partial \alpha_{i-1}}{\partial x_k} \varphi_k$, $i = 2, \cdots, n$. }} \\ \\ \\

    \bottomrule
    \end{tabular}}
\end{table}

Given the existence of input saturation, in the final step, the following dynamic variable $\xi$ is incorporated: 
\begin{align}\label{xi}
    \dot{\xi} = -\lambda\xi + [h(v) - v],   \quad \xi(0) = 0,
\end{align}
where $\lambda > 0$ is a design constant.
Similarly, the design process, which considers the disturbances and input saturation, is summarized in \hyperref[table2]{Table II}.
The stability analysis of the closed-loop system under the proposed adaptive controller is summarized in the following theorem.

\emph{Theorem 3:} Consider the nonlinear system (\ref{system}) under the input saturation (\ref{Sat_u}).
The PF is given by (\ref{PPF}), while the decay rate of the PF is specified as (\ref{d_delta}).
The control law with the parameter update law is designed in \hyperref[table2]{Table II}.
Under Assumptions \hyperref[assumption1]{1-3}, it is ensured that:
1) all signals in the closed-loop systems are ultimately uniformly bounded, and
2) the PF (\ref{PPF}) is never violated by $|e(t)|$.

\emph{Proof:}
Following the backstepping design in \hyperref[table2]{Table II}, we can obtain the derivative of $V_n$ as follows:
\begin{align}\label{dV_n-2}
    \dot{V}_n \leq&\ -\sum_{k=1}^{n} c_k z^2_k + \sum_{k=1}^{n} \frac{(n+1-k)\bar{d}^2_k}{4} + \frac{\bar{g}^2}{4} \nonumber \\
    \leq &\ -\rho V_n + b,
\end{align}
where $\rho = \text{min}\{c_1, \cdots, c_n, 1\} > 0$, $b = \sum_{k=1}^{n} \frac{(n+1-k)\bar{d}^2_k}{4} + \frac{\bar{g}^2}{4} + \frac{2\theta^2_M}{\lambda_{\text{min}}(\Gamma)} < \infty$.
Solving the inequality (\ref{dV_n-2}) gives
\begin{align}\label{V_n-2}
    V_n(t) \leq V_n(0)\text{exp}^{-\rho t} + \frac{b}{\rho}(1 - \text{exp}^{-\rho t}).
\end{align}
Hence, $V_n(t)$ is bounded for all $t \geq 0$.
Then, it can be checked that all signals within the closed-loop systems are ultimately uniformly bounded.

Moreover, the boundedness of $s$ implies that the defined PF (\ref{PPF}), which is independent of initial conditions and has a self-tuning decay rate (\ref{d_delta}), will never be violated by $|e(t)|$, i.e.,
\begin{align}
    I(-\Psi(t))< e < I (\Psi(t)), \quad \text{for} \ \forall t \geq 0.
\end{align}
The proof is completed.
$\hfill \blacksquare$

\begin{figure}[tp]\label{fig2}
\centering
\setlength{\belowcaptionskip}{0mm}
\includegraphics[scale=0.64]{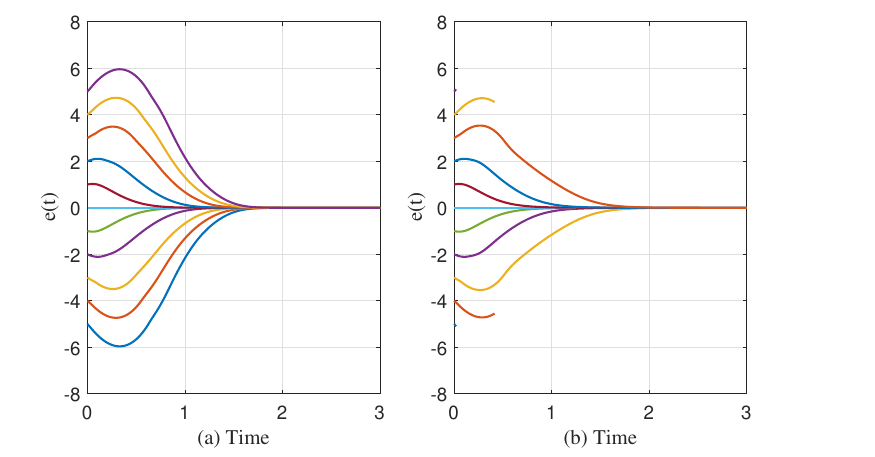}
\captionsetup{font={small}}
\caption{The tracking error in (a) this work, and (b) work \cite{zhao2021adaptive}. \qquad\qquad\qquad}
\setlength{\abovecaptionskip}{1cm}   
\end{figure}

To test the proposed method, we initially consider an uncertain dynamic system borrowed from \cite{zhao2021adaptive}:
\begin{equation}\label{system_2}
    \left\{
        \begin{aligned}
            & \dot{x}_1 = x_2 + \theta^T \varphi_1, \\
            & \dot{x}_2 = u,\\
            &y = {{x}_{1}},
        \end{aligned}
    \right.
\end{equation}
with the input saturation limit $\bar{u} = 20$, and $\theta = 1$, $\theta_M = 2$, $\varphi_1 = x_1$.
To validate the superiority of this work, we compare it with \cite{zhao2021adaptive} and set $x_2(0) = 0$.
The desired signal is given as $y_d(t) = 0$.
The control parameters are selected as $c_1 = 5$, $c_2 = 10$, $\Gamma = 1$, $\Psi_\infty = 0.05$, and $\lambda = 5$.
The initial estimate of the parameter is $\hat{\theta}(0) = 0$.
Additionally, we set the parameter $\delta$ in \cite{zhao2021adaptive} as $\delta = 0.05$, and in our work, we set $\varrho_{1} = 2$, $\varrho_2 = 10$, and $\varepsilon_0 = 0$.
Finally, the simulation results are presented in \hyperref[fig2]{Fig. 2}.
From \hyperref[fig2]{Fig. 2}, it can be observed that.
Firstly, for a chosen exponential convergence rate $\delta$, the method proposed in \cite{zhao2021adaptive} is not applicable to all initial states when input saturation is present, although the systems corresponding to these initial states still satisfy \hyperref[assumption3]{Assumption 3}.
Secondly, for the method proposed in \cite{zhao2021adaptive}, the restriction on the initial states of the system can be relaxed by selecting a smaller $\delta$.
However, it is worth noting that a smaller $\delta$ implies a slower decay rate for the entire convergence process of the tracking error, leading to a longer convergence time, as evident from the comparison in the left and right subplots of \hyperref[fig2]{Fig. 2}.
Therefore, it can be concluded that the proposed method in this work not only eliminates the need for pre-setting the decay rate of the PF, but also achieves better control performance.
Specifically, on one hand, compared to previous PPC methods based on fixed exponential decay rates, our method allows for looser initial error values.
On the other hand, for the same initial error values, our method requires a shorter convergence time.

\section{Numerical Simulations on MSD system}

\begin{figure}[tp]\label{fig3}
\centering
\setlength{\belowcaptionskip}{0mm}
\includegraphics[scale=0.5]{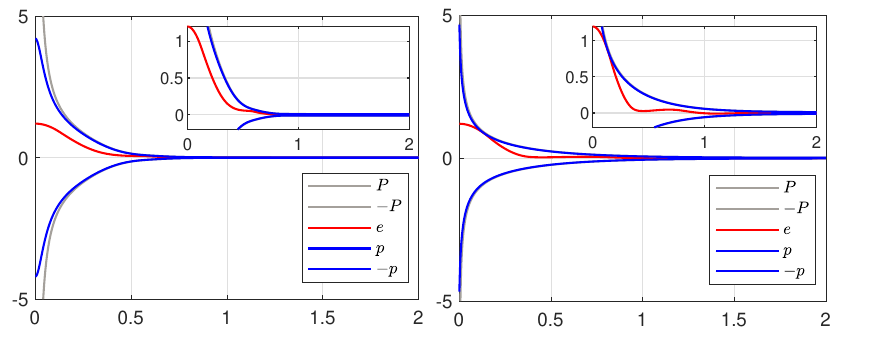}
\includegraphics[scale=0.5]{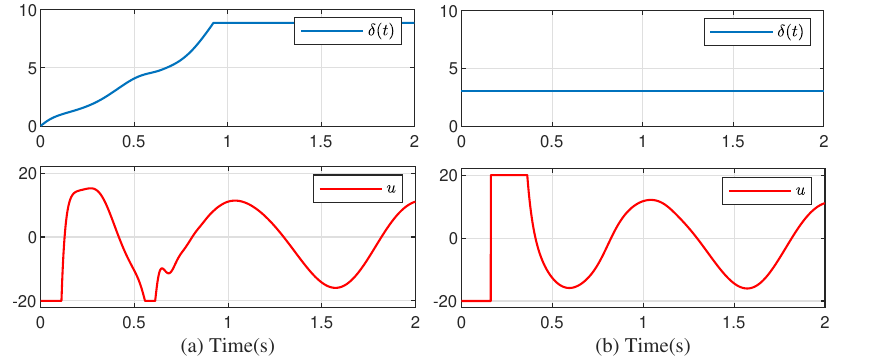}
\captionsetup{font={small}}
\caption{The control performance corresponding to $x_1(0) = 0.6$ m; (a) this work, (b) work \cite{zhao2021adaptive}.}
\setlength{\abovecaptionskip}{1cm}   
\end{figure}

\begin{figure}[tp]\label{fig4}
\centering
\setlength{\belowcaptionskip}{0mm}
\includegraphics[scale=0.51]{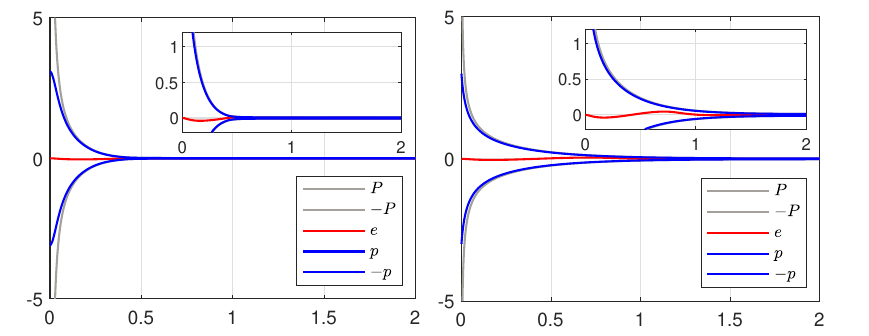}
\includegraphics[scale=0.51]{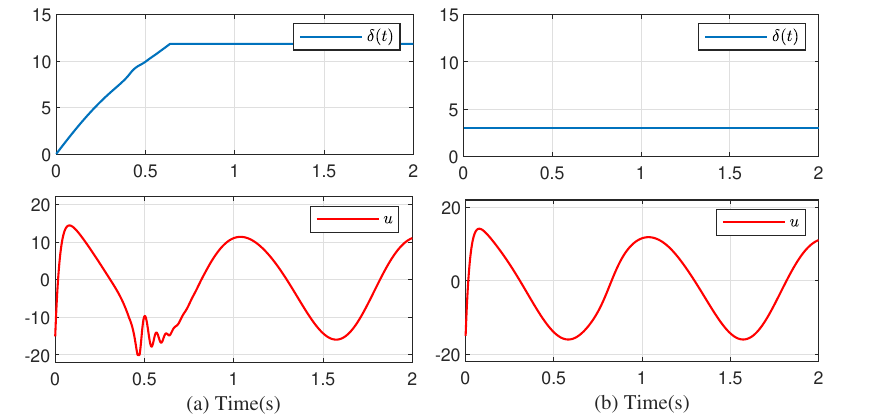}
\captionsetup{font={small}}
\caption{The control performance corresponding to $x_1(0) = -0.6$ m; (a) this work, (b) work \cite{zhao2021adaptive}.}
\setlength{\abovecaptionskip}{1cm}   
\end{figure}

In this section, we further consider the mass-spring-damper (MSD) system utilized in \cite{wen2011robust} as follows:
\begin{equation}\label{system_s}
    \left\{
        \begin{aligned}
            & \dot{x}_1 = x_2, \\
            & \dot{x}_2 = u(v) - \frac{k}{m}x_1 - \frac{c}{m}x_2 + d(t),\\
            &y = {{x}_{1}},
        \end{aligned}
    \right.
\end{equation}
where $x_1$ and $x_2$ represent the position and velocity of the object, respectively;
$m$ is the mass of the object, $k$ is the stiffness constant of the spring and $c$ is the damping, and the external disturbance $d(t) = \sin(4\pi t)$.
The input saturation limit is $20$ N.
The parameters $m = 1$ kg, $c = 2$ N$\cdot$s/m, and $k = 8$ N/m are unknown, but the norm of the vector $\theta = [-k/m, -c/m]^T$ is known to have an upper bound of $\theta_M = 10$.
The desired trajectory is specified as $y_d = -0.4\cos(2\pi t)-0.2$ [m].
Similarly, to illustrate the superiority of this work, we compare it with work \cite{zhao2021adaptive}.
The control parameters are selected as $c_1 = 20$, $c_2 = 25$, $\Gamma = [1,0;0,1]$, $\Psi_\infty = 0.01$, and $\lambda = 20$.
The initial conditions for the first and second instances are chosen as $x_1(0) = \pm 0.6$ m and $x_2(0) = 0$ m/s, respectively.
The initial estimate of the parameter is $\hat{\theta}(0) = [0, 0]^T$.
Additionally, we set the parameter $\delta$ in \cite{zhao2021adaptive} as $\delta = 3$, and in our work, we set $\varrho_1 = 25$ and $\varrho_2 = 10$.
Finally, the simulation results for the system tracking error $e(t)$, the decay rate of the PF $\delta(t)$, and the control signal $u(t)$ for the first and second instances are presented in \hyperref[fig3]{Fig. 3} and \hyperref[fig4]{Fig. 4}, respectively.
By comparing \hyperref[fig3]{Fig. 3} and \hyperref[fig4]{Fig. 4}, we can conclude that although previous PPC methods with a fixed decay rate can artificially select a very small $\delta$ to ensure that the maximum initial error encountered in practice still satisfies the PF constraint under specific input saturation limits, it actually sacrifices the convergence time corresponding to small initial errors.
While the proposed method in this work can effectively avoid this issue.

\section{Conclusions}
This work proposes a self-tuning performance function based controller for a class of nonlinear systems with uncertain parameters and input saturation constraints.
By introducing a performance index function as a baseline, the decay rate of the performance function can be adaptively adjusted based on certain criteria, allowing it to avoid violations of the performance function while enhancing the tracking performance of the system.
Overall, this research contributes to improving system performance, specifically addressing the challenges posed by actuator saturation and performance constraints.
In the further research, it is warranted to extend the results to cover more general classes of systems.
\end{document}